\documentstyle[12pt]{article}
\def\beq{\begin{equation}}   \def\eeq{\end{equation}}
\def\be{\begin{equation}}   \def\ee{\end{equation}}
\def\bea{\begin{eqnarray}}   \def\eea{\end{eqnarray}}

\begin{document}

\begin{flushright}
UND-HEP-02-BIG\hspace*{.2em}03\\
hep-ph/0203157
\end{flushright}
\vspace{.3cm}
\begin{center} 
\Large 
{Looking from the East at an Elephant Trotting West 
\footnote{With due apologies to Sir Charles Laughton. }:   
Direct $CP$ Violation in $B^0$ Decays} 
\\
\end{center}
\vspace*{.3cm}
\begin{center} 
{\Large
 I. I. Bigi \\
\vspace{.4cm}
{\normalsize 
{\it Physics Dept.,
Univ. of Notre Dame du
Lac, Notre Dame, IN 46556, U.S.A.} }   
} \\
\vspace{.3cm}
e-mail address:\\
{\it  bigi.1@nd.edu} 
\vspace*{1.4cm}

{\Large{\bf Abstract}}\\
\end{center}

\noindent 
With the time dependant CP asymmetry in $B_d(t) \to \psi K_S$ well
measured, the most powerful method to search for {\em direct} 
CP violation in $B_d$ decays is to analyze whether the 
CP asymmetry of other transitions  
like $B_d(t) \to \pi ^+ \pi ^-, \eta ^{\prime}K_S$ etc. 
can be described
by a 
$sin$ term taken from $B_d(t) \to \psi K_S$ and without a 
$cos$ term. The failure of such a constrained fit would establish direct 
CP violation (even if no CP asymmetry were observed in 
these other modes).

\vspace*{.2cm}
\vfill
\noindent
\vskip 5mm



\noindent 
{\bf Prologos}
\vspace*{0.2cm}

The occurrance of large $CP$ violation in $B_d \to \psi K_S$  
predicted in 1981 \cite{BS} has been  
established in 2001 
\cite{BABAR1,BELLE1}. BABAR and BELLE  have just now presented updates: 
\bea 
{\rm sin}(2\phi _1) &=& 0.82 \pm 0.12 \pm 0.05 \; \; \; 
{\rm BELLE} \; \cite{BELLE2} 
\label{phi1}
\\
{\rm sin}(2\beta) &=& 0.75 \pm 0.09 \pm 0.04 \; \; \; 
{\rm BABAR} \; \cite{BABAR2}
\label{beta}
\eea 
These findings are 
of a paradigmatic nature in several respects even beyond the 
obvious one that they  
constitute the first observation of CP violation outside 
the $K_L$ system: (i) The effect being truly large `de-mystifies' 
CP violation: for it tells us that if CP invariance can be 
broken, that violation is not
intrinsically small \cite{DENT}. 
(ii)  While the observed asymmetry could
still contain sizeable contributions from New Physics, it conforms to the
CKM prediction in a completely unforced way.
It  behooves us now to refer to the CKM description of CP violation as a 
{\em theory} rather than giving it the somewhat patronizing label
 {\em ansatz}.

\vspace*{0.5cm} 

\noindent 
{\bf Parabasis} 
\vspace*{0.2cm}

Despite the success of the CKM theory so far, it would be 
premature to accept even the overall pattern of its predictions and
consider  only deviations from it as noteworthy. In particular there is 
another paradigmatic issue to be raised concerning the 
classification of the underlying dynamics: is there  
also {\em direct} CP violation, i.e. CP violation in 
$\Delta B =1$ transitions? The CKM theory definitely 
does not  produce a superweak scenario for $B$ decays -- 
alas, one still has to verify it.

Observing a CP asymmetry in charged $B$ decays would establish it 
unequivocally. The situation concerning neutral $B$ decays is  
more subtle. Consider 
\beq 
e^+e^- \to \Upsilon (4S) \to B_d \bar B_d \to f_{tag} f_{CP} 
\eeq
where $f_{tag}$ denotes a final state tagging the flavour of one 
beauty meson and $f_{CP}$ a CP eigenstate, into which the other beauty 
meson decays and which is reconstructed. Then one finds for the 
decay rate $R_+$ [$R_-$] where the tagging decay is that of a $B_d$ 
[$\bar B_d$]: 
$$  
R_{\pm}(\Delta t) \propto e^{-|\Delta t|\Gamma (B_d)} \times 
\left[ 
1 \pm S {\rm sin} (\Delta m_d \Delta t) 
\pm C {\rm cos} (\Delta m_d \Delta t)
\right] 
$$ 
\beq 
S \equiv \frac{2{\rm Im}(q/p)\bar \rho (f_{CP})}
{1+ |\frac{q}{p}\bar \rho (f_{CP})|^2} \; , \; 
C \equiv - \frac{1- |\frac{q}{p}\bar \rho (f_{CP})|^2}
{1+ |\frac{q}{p}\bar \rho (f_{CP})|^2} 
\label{DEFSC}
\eeq 
with $\Delta t$ denoting the difference in (proper) time of the 
two decays [$\Delta t \equiv t_{CP} - t_{tag}$],  
$\bar \rho (f_{CP})$ the ratio between the two decay {\em amplitudes}  
[$\bar \rho (f_{CP}) \equiv T(\bar B_d \to f_{CP})/T(B_d \to
f_{CP})$] and $q/p$ coming from the diagonalization of the 
$B_d - \bar B_d$ mass matrix. The
asymmetry  then reads: 
\beq  
\frac{R_+(\Delta t) - R_-(\Delta t)}{R_+(\Delta t) + R_-(\Delta t)} 
= S  {\rm sin} (\Delta m_d \Delta t) + 
C {\rm cos} (\Delta m_d \Delta t)  
\eeq
 Observing a $cos$ term in the time evolution establishes direct CP
violation, since  in that case $|T(\bar B \to f_{CP})|^2 \neq 
|T( B \to f_{CP})|^2$ --
an obvious and well-known  statement. 
For this to happen, two transition amplitudes with different weak 
and strong phases have to contribute; this situation is realized 
due to the intervention of Penguin operators.

However -- as is known as well \cite{BIGI}, though less appreciated -- 
direct CP violation can manifest itself also in a different 
way. The quantity 
$(q/p)\bar \rho (f_{CP})$ reflects $\Delta B =2$ and $\Delta B = 1$ 
dynamics in the first and second factors, respectively;  
$(q/p)$ and $\bar \rho (f_{CP})$ separately 
depend on the phase convention $\xi$ adopted for the definition of 
antiparticles: $|\bar B_d \rangle \equiv e^{i\xi}{\bf CP} |B_d\rangle$.  
Their product, however, is phase convention independant and thus an
observable. As long as CP violation is studied in a single channel 
$B_d (t) \to f_{CP}$, one cannot draw an empirical
distinction  between  classifying a signal as `superweak' or
`direct' from measuring the $sin$ term; for a change in the  phase
convention $\xi \to \xi +\Delta \xi$ can make either
$q/p$ or
$\bar \rho (f_{CP})$  real. 
A superweak scenario of CP violation is thus best defined as one where 
there exists a convention for the phase $\xi$ s.t. all 
$\Delta B=1$ decay amplitudes are real. 
However once one studies the time evolution
in two  different channels 
\beq 
B_d (t) \to f_{CP} \; \; vs. \; \; B_d (t) \to \tilde f_{CP}
\eeq
an empirical distinction can be drawn: in a {\em superweak} 
scenario the $cos$ term is effectively absent 
since $|(q/p)\bar \rho (f_{CP})|^2 =  
|(q/p)\bar \rho (\tilde f_{CP})|^2 = |q/p|^2 \simeq 1$ with 
$|q/p| = 0.998 \pm 0.006 \pm 0.007$ experimentally 
\cite{AUBERT} 
\footnote{Theoretically one expects 
$|1 - |q/p|^2| \sim {\cal O}(10^{-3})$.} 
; likewise 
the $sin$ term is then of equal size for 
both modes: 
\beq 
S(f_{CP}) = 
\frac{2}{1+ \left| \frac{q}{p}\right| ^2} 
{\rm Im}\frac{q}{p}\bar \rho (f_{CP}) = 
\eta _f \eta _{\tilde f} S(\tilde f_{CP}) 
\; , 
\eeq  
where $\eta _f$ and $\tilde \eta _{\tilde f}$ denote the CP parities 
of $f_{CP}$ and $\tilde f_{CP}$, respectively. 
Any deviation from this simple pattern reveals 
{\em direct} CP violation! One should also note that if the 
two transitions $B_d \to f_{CP}$ and $B_d \to \tilde f_{CP}$ 
are each described by a single amplitude or if there are 
no significant final state interactions in those two modes, then 
one has $|\bar \rho (f_{CP})|^2 = 1 = |\bar \rho (\tilde f_{CP})|^2$.  
Yet Im$(q/p)\bar \rho (f_{CP}) \neq$ Im$(q/p)\bar \rho (\tilde f_{CP})$ 
can hold due to different {\em weak} phases in the $\Delta B =1$ 
amplitudes for $B_d \to f_{CP}$ and $B_d \to \tilde f_{CP}$. I.e., such 
direct CP violation could not manifest itself through a 
$cos$ term, only through a difference in the $sin$ terms.

Such a method becomes quite powerful now, since the well measured 
asymmetry in $B\to \psi K_S$, Eqs.(\ref{phi1},\ref{beta}), provides an
excellent calibrator: the $sin$ term has been well measured and a fairly
tight bound placed on the $cos$ term. I.e., one considers other 
$B_d \to \tilde f_{CP}$ 
decays into CP eigenstates and analyzes whether the time evolution of 
the rate can adequately be described without a $cos$ term and with a 
$sin$ 
term that has the same [opposite] coefficient as in $B_d \to \psi K_S$ 
if $\tilde f_{CP}$ is CP odd [even].

BELLE and $BABAR$ have presented data on two such additional channels: 
\begin{itemize}
\item 
BELLE describes the time evolution of the asymmetry in the decay 
$B_d \to \eta ^{\prime} K_S$ with just a $sin$ term \cite{BELLE2}:
\beq 
{\rm Im}\frac{q}{p} \bar \rho (\eta ^{\prime}K_S) = 
0.29 ^{+0.53}_{-0.54} \pm 0.07
\eeq
This result is consistent with no asymmetry, which would imply large 
{\em direct} CP violation to make it consistent with the 
findings in $B_d \to \psi K_S$. At the same time it falls 
within  one sigma of a superweak CP violation 
(both $\eta ^{\prime} K_S$ and $\psi K_S$ are CP odd). 
\item 
BELLE fits the $B_d (t) \to \pi ^+ \pi ^-$ evolution using a $sin$ as 
well as $cos$ term with coefficients \cite{BELLE2}
\bea
S &=& - 1.21 ^{+0.38+0.16}_{-0.27-0.13} \\ 
C &=& + 0.94 ^{+0.25}_{-0.31} \pm 0.09 \; ; 
\eea
$BABAR$ on the other hand finds \cite{BABAR3} 
\footnote{With the definition for $C$ given in Eq.(\ref{DEFSC}) 
$C$ has the opposite sign from the $BABAR$ convention.} : 
\bea
S &=& - 0.01 \pm 0.37 \pm 0.07 \\ 
C &=& + 0.02 \pm 0.29 \pm 0.07 \; .  
\eea
Those two sets of numbers convey the sense of two different
messages: the BELLE  analysis presents tantalizing evidence for a CP
asymmetry in a second  class of $B$ decays that furthermore has a large
direct component, whereas the $BABAR$ study shows no evidence for a CP
asymmetry.  Future measurements will clarify the experimental picture.

The first step towards properly 
{\em interpreting} the data is to analyze 
whether the time evolution of 
$B_d (t) \to \pi ^+ \pi ^-$ contains a {\em qualitatively new}  
feature beyond what is seen in $B_d (t) \to \psi K_S$, 
i.e. whether there is direct CP violation. 
This question has to be answered experimentally 
irrespective of the KM theory predicting such an effect! 
This is most
efficiently  achieved by studying whether the CP asymmetry in 
$B_d (t) \to \pi ^+ \pi ^-$ can be described with $C=0$ and 
$S_{\pi\pi} = - S_{\psi K_S}$ ($\pi \pi$ and $\psi K_S$ 
carry opposite CP parity), see Eqs.(\ref{phi1},\ref{beta}), i.e. 
{\em without} any free fit parameter! What drives 
$B_d (t) \to \psi K_S$ has to contribute to 
$B_d (t) \to \pi ^+ \pi ^-$ as well. 
{\em Before  
such a  `minimal' scenario 
has not been ruled out, it is premature to undertake 
fits  where $S$ and $C$ are allowed to float!} Only after 
direct CP violation has been established due to the failure 
of a superweak fit (or quantitative theoretical 
control has been established over the impact of Penguin operators)  
should one concentrate on fitting the data with $S$ and $C$ 
not a priori fixed. The goal there is to separate the two variants of
direct CP violation, namely  Im$\frac{q}{p}\bar \rho (\psi K_S) \neq -$ 
Im$\frac{q}{p}\bar \rho (\pi \pi)$ and 
$|\bar \rho (\pi \pi )|^2 \neq 1$,  
and ultimately also of isolating the Penguin contribution.
  
One should note that a fit to $B_d(t) \to \pi ^+\pi^-$ 
yielding $S\simeq 0 \simeq C$ as suggested by the $BABAR$ 
findings -- viewed
together with the  data on $B_d(t) \to \psi K_S$ -- would reveal the
intervention of  large {\em direct} CP violation {\em without} exhibiting 
any CP asymmetry!
 
\end{itemize}

{\em In summary}: The main point of this short note is to emphasize 
that one should and can 
empirically establish the categorical feature whether direct CP
violation occurs in $B_d$ decays. 
This is best achieved by analyzing whether the time evolution of a 
CP asymmetry in $B_d(t) \to \tilde f_{CP}$, 
$\tilde f_{CP} \neq \psi K_S$ can be described {\em without} a 
$cos$ term and with a $sin$ term of equal size as in 
$B_d(t) \to \psi K_S$. 
Once it has been decided that such a description without a free 
fit parameter 
fails,  one can address 
the more challenging task to extract the coefficients of the 
$sin$ and $cos$ terms in an unconstrained fit with the ultimate 
goal of determining 
the size of the Penguin contribution, phases etc.

\vspace*{0.5cm} 

\noindent 
{\bf Epilogos}
\vspace*{0.2cm}

The great actor Sir Charles Laughton once referred to his face as 
looking like the east side of an elephant walking west. 
The point of analogy here differs from this imagery, though: 
Evaluating the beauty of an elephant is a rather complex task, yet 
spotting him from any direction 
should be quite straightforward.

\vspace*{.2cm}

{\bf Acknowledgements:}~~
This work has been supported by the 
National Science 
Foundation under grant number PHY00-87419.


\end{document}